\documentclass[prd,showpacs,superscriptaddress,twocolumn,
floatfix,preprintnumbers,altaffilletter]{revtex4}
\usepackage{graphicx,url,amssymb,amsmath,longtable,rotating,color,units, 
wasysym,subfigure,epsfig,multirow}

\usepackage{epstopdf}

\begin{document}
\pacs{95.75.-z,04.30.-w,07.05.Bx}

\title{Chasing 5-sigma: Prospects for searches for long-duration gravitational-waves without time slides}


\author{Michael Coughlin}
\email{coughlin@physics.harvard.edu}
\affiliation{Department of Physics, Harvard University, Cambridge, MA 02138, USA}

\author{Patrick~Meyers}
\affiliation{School of Physics and Astronomy, University of Minnesota, Minneapolis, MN 55455, USA}

\author{Shivaraj~Kandhasamy}
\affiliation{Physics and Astronomy, University of Mississippi, University, MS 38677-1848, USA}

\author{Eric~Thrane}
\affiliation{School of Physics and Astronomy, Monash University, Clayton, Victoria 3800, Australia}

\author{N Christensen}
\affiliation{Physics and Astronomy, Carleton College, Northfield, Minnesota 55057, USA}

\date{\today}

\begin{abstract}

  The detection of unmodeled gravitational-wave bursts by ground-based interferometric gravitational-wave detectors is a major goal for the advanced detector era.
  These searches are commonly cast as pattern recognition problems, where the goal is to identify statistically significant clusters in spectrograms of strain power when the precise signal morphology is unknown.
  In previous work, we have introduced a clustering algorithm referred to as ``seedless clustering,'' and shown that it is a powerful tool for detecting weak long-lived ($\sim$$10$--$\unit[1000]{s}$) signals in background.  
  However, as the algorithm is currently conceived, in order to carry out an all-sky search on a $\approx$ year of data, significant computational resources may be required in order to carry out background estimation.  
  Alternatively, some of the sensitivity of the search must be sacrificed to control computational costs.
  The sensitivity of the algorithm is limited by the amount of computing resources due to the requirement of performing background studies to assign significance in gravitational-wave searches.
  In this paper, we present an analytic method for estimating the background generated by the seedless clustering algorithm and compare the performance to both Monte Carlo Gaussian noise and time-shifted gravitational-wave data from a week of LIGO's 5th Science Run.
  We demonstrate qualitative agreement between the model and measured distributions and argue that the approximation will be useful to supplement conventional background estimation techniques for advanced detector searches for long-duration gravitational-wave transients.
\end{abstract}

\maketitle

\section{Introduction}\label{intro}
Second-generation gravitational-wave detectors such as Advanced LIGO~\cite{aligo} and Advanced Virgo~\cite{avirgo} will be coming online in the coming months and years. 
Some searches for gravitational-wave bursts seek to detect signals lasting less than $\approx$1\,s \cite{PhysRevD.85.122007}. 
Other searches target signals lasting $\sim$$10$--$\unit[1000]{s}$. 
Compact binary coalescences of black holes (and/or neutron stars) are one example of long-lived gravitational-wave sources \cite{S6Highmass,s6lowmass,rates}.
Theoretically uncertain models exist for more exotic sources of long-lived transients, including emission from rotational instabilities in protoneutron stars~\cite{pirothrane12,piro:11,piro:07,corsi} and black-hole accretion disk instabilities~\cite{kiuchi,vanputten:01,vanputten:08}.
When a matched filter search is not possible, searches for long-lived bursts~\cite{lgrb,stamp,stochtrack,stochsky,stochtrack_cbc,stochtrack_ecbc} can be employed.

Searches for compact binary coalescences rely on performing time-slides of single detector ``triggers,'' generated by performing a matched filter (in the compact binary coalescence case) on single-detector time series data.
Burst searches instead use clustering algorithms on single detector time-frequency maps \cite{box} or multi-detector coherent maps \cite{X-Pipeline,CoherentWaveBurst,stamp}.
Calculating the coherent statistic takes significant computational resources.
Because detector noise is generally non-Gaussian, it is difficult to know if an event in one detector is signal or noise. 
For this reason, multiple detectors are required to perform gravitational-wave searches. 
To estimate background, these searches time-shift the data of each detector with respect to the other(s), by some unphysical delay which is larger than the light travel time between the detectors. The coherent statistics are then computed between the time-shifted data in the same way as the original search algorithm, and in this way, false alarm rates can be estimated. 
Current searches use thousands (or more) of time-slides \cite{PhysRevD.85.122007,s6lowmass,s6grb}. The main limitation on the number of time-slides that can be performed is limited computational resources, although the short-duration coherent burst pipelines have been tested in the ten-thousand time-slides regime and are mvoing towards the hundred-thousands while the compact binary matched filtering pipelines have successfully generated $5 \sigma$ background distributions. Th difficulty in reaching these levels is due to computationally intensive calculations like the matched filter used in compact binary coalescence \cite{PhysRevD.71.062001}, calculation of the coherent SNR \cite{0264-9381-22-18-S46} used in burst searches, and potentially seedless clustering~\cite{stochtrack,stochsky,stochtrack_cbc,stochtrack_ecbc} in a long-duration transient search.

Was et al.\,demonstrated the limitation of using time slides to perform background estimation in the single-detector trigger case (this is not generally a problem for coherent analyses) \cite{0264-9381-27-1-015005,0264-9381-27-19-194014}. They showed that the precision on the background estimation using time slides
of trigger streams is in fact limited and that the variance associated with their use saturates at some point. 
The computational limitations and the potential problems with time-slides motivate a search for potential alternative forms of background estimation in gravitational-wave searches. Gravitational-wave searches for isotropic stochastic gravitational wave backgrounds \cite{S5Stochastic,S6Stochastic} and directional searches towards Sco X-1, the galactic center, and SN1987A \cite{PhysRevLett.107.271102} have assumed that the detection statistic is normally distributed with a known mean and variance that can be calculated from first principles when performing the searches. These searches sum up data from long stretches of time, and combined with the use of long time segments (60\,s) and Gaussianity cuts, these statistics are Gaussian by the Central Limit Theorem. This has the significant computational cost-saving benefit of not requiring time slides to perfom the search, although limited time-shift analyses are used as sanity checks and to ensure that particularly non-Gaussian frequency bins can be removed from the analysis. 

Some searches for long-duration gravitational-wave bursts use the same cross-correlation technique as stochastic searches \cite{stamp}, although other methods exist \cite{X-Pipeline,CoherentWaveBurst}. 
They utilize cross-power spectrograms, computed from the cross-correlation of two gravitational-wave detectors, and use pattern recognition algorithms to search for clusters of excess strain cross-power \cite{stamp}.  
One algorithm used to search for long-duration gravitational waves is ``seedless clustering,'' which integrate along many different paths in spectrograms. 
This algorithm is sensitive to signals that can be well-approximated by parameterized curves, and the advantage of seedless clustering is most pronounced for long and weak signals \cite{stochtrack,stochsky,stochtrack_cbc,stochtrack_ecbc}. 
We have previously shown how seedless clustering algorithms can be used to significantly enhance the sensitivity of searches for signals of this type~\cite{stochtrack}. 
Although seedless clustering algorithms are ``embarrassingly parallel''~\cite{parallel}, and therefore computations can be performed on graphical processor units, seedless clustering searches are still limited by computation of the noise background.

Cannon et al.\,\cite{CaHa2013} recently proposed a method to estimate the false alarm probability of compact binary coalescences without time-shifts. They rely on the approximation of compact binary events as a Poisson process. This in particular allows for a statistical detection of a population of events, which could be collectively more significant than the single most significant event alone. The method proposed in our paper is similar in that we measure events based on the data and then use a statistical approximation to the distribution of the measured tracks to make approximations to the noise background. There are also a number of notable differences. Because long-duration transient gravitational waves are typically searched for using a coherent combination of detector data, the trigger distributions no longer obey Poisson statistics. Instead, we will exploit the fact that seedless clustering sums many approximately statistically independent pixels to use Gaussian statistics to estimate the background.
In this paper, we demonstrate a semi-analytical approximation to the seedless clustering output from cross-correlation spectrograms.
One potential criticism of the analysis that follows is the fact that we compare the approximation with data from time-shifted analyses out to $\approx 3 \sigma$, not to the $5 \sigma$ distributions we present at the end of the paper.  
This is a ``chicken-or-the-egg'' problem in the following sense: in order to truly verify the approximation we use, it would be necessary to perform $5 \sigma$ worth of time-shifts. 
This calculation is currently very difficult to do computationally, and of course, if we could perform $5 \sigma$ worth of time-shifts, we would not need an approximation in the first place.
Moreover, as we perform the analysis using a relatively clean week-long stretch of data, different sets of data could result in different results.
Therefore, we consider the analysis that follows as a first test for the feasibility of an approximate method. 
As argued above, we expect the background distributions to be better behaved in long-duration analyses than in short-duration searches, and therefore perhaps less susceptible to significant deviations from empirical distributions.
In the future, we can use distributions generated by future analyses that perform more time slides and over longer periods to compare against the approximation to test its utility.
Therefore, although time-slides are likely required to create confidence in a detection due to the problem just described, we now summarize several reasons why it is useful to consider alternative significance-estimation strategies.

{\em Algorithm Verification.}
The semi-analytic method provides a verification for the pipeline in multiple ways. 
In the case where data-quality work is being performed correctly, in general, the data should be generally well-approximated by Gaussian noise, outside of some data transients which pass the data quality cuts. Therefore, background estimation should approximately follow the distribution if it is assumed that the data is Gaussian.
Also, this provides a sanity check that the algorithm performs as expected on the data.
By performing a limited number of time-slides or running on Gaussian noise, it should be clear that the model for the algorithm is correct, which can provide confidence that the algorithm is performing as expected (or not).

{\em Sensitivity to waveform models.}
There are a number of papers contained in the literature about the sensitivity of gravitational-wave detectors to long-duration gravitational waves \cite{stochtrack,stochsky,pirothrane12,stochtrack_cbc,stochtrack_ecbc}. In general, the sensitivity studies have been performed by running the analysis on 1,000 $ft$-maps to reach a FAP of 0.1\%, and the sensitivity to various waveform models are computed relative to this number. 
For a year of data, assuming $ft$-maps of $\unit[250]{s}$ with 50\% overlap, and a desirable FAP of $\approx 3\,\sigma$ or $\text{FAP}=0.27\%$, there will be more than $10^8$ maps analyzed. 
Before any analysis, either a search for gravitational-waves or a waveform sensitivity study, is performed, it is desirable to be able to estimate the background quickly.  
This estimate informs expectations of potential results as well as how to setup the analysis.
Using the method described in this paper, we can analytically compute what we expect a threshold based on this number of maps, without needing to run nearly so many time-slides.

{\em Event follow-up and electromagnetic alerts.}
There are preparations for joint electromagnetic and gravitational-wave observations in the advanced detector era \cite{SiPr2014}. Low-latency gravitational-wave searches are aiming for run-times $\leq \unit[1]{\text{min}}$. For significance estimates on this time-scale, rapid background estimation techniques are required.
The method described in this paper is able to give an approximate FAR for any event on this time-scale. 
In the case where there is eventually interest in joint electromagnetic and gravitational-wave observations for generic long-duration transients, this method may be useful for making that happen.


The remainder of the paper is organized as follows.
In Sec.~\ref{formalism}, we describe the formalism of an all-sky transient search and seedless clustering.
In Sec.~\ref{results}, we present the results of a Monte Carlo and time-shifted study comparing the semi-analytical model to seedless clustering.
In Sec.~\ref{conclusions}, we discuss our conclusions and suggest directions for future research.

\section{Formalism}\label{formalism}

We use the cross-correlation of two GW strain channels from spatially separated detectors to perform searches for long-duration GW transients.
We construct $ft$-maps of cross-power signal-to-noise ratio.
We divide detector strain time series into segments and compute Fourier transforms of the segments to create the pixels, which we denote as $\tilde{s}_I(t;f)$, where we take strain data from detector $I$ for the segment with a mid-time of $t$.
Following \cite{stochtrack,stochsky,stochtrack_cbc,stochtrack_ecbc}, the segments are $50\%$-overlapping and Hann-windowed with duration of $\unit[1]{s}$ and a frequency resolution of $\unit[1]{Hz}$.

The expression for the cross-power signal-to-noise ratio is as follows \cite{stamp}:
\begin{equation}\label{eq:rho}
  \rho(t;f|\hat\Omega) = \frac{2 \sqrt{2}}{\cal N} \text{Re}\left[
    e^{2\pi i f \Delta\vec{x}\cdot\hat\Omega/c}
    \frac{\tilde{s}_I^*(t;f) \tilde{s}_J(t;f)}{\sqrt{P'_I(t;f) P'_J(t;f)}}
    \right] .
\end{equation}
where $\Delta\vec{x}$ is a vector describing the relative displacement of the two detectors, $\hat\Omega$ is the direction of the GW source, and $c$ is the speed of light. The time delay between the two detectors, which is a direction-dependent phase factor, is in the $e^{2\pi i f \Delta\vec{x}\cdot\hat\Omega/c}$ term.
$P'_I(t;f)$ and $P'_J(t;f)$ are the auto-power spectral densities for detectors $I$ and $J$ in the segments neighboring $t$. $\cal N$ is a FFT normalization factor, $L \times Fs$, where $L$ is the length of data in seconds and $Fs$ is the sampling frequency.
Additional details can be found in~\cite{stamp,stochtrack,stochsky,stochtrack_cbc,stochtrack_ecbc}

We write the total signal-to-noise ratio for a cluster of pixels as a sum of over $\rho(t;f|\hat\Omega)$:
\begin{equation}\label{eq:sum}
  \text{SNR}_\text{tot} \equiv
  \frac{1}{N^{1/2}}
  \sum_{\left\{t;f\right\}\in\Gamma} \rho(t;f|\hat\Omega) ,
\end{equation}
where $N$ is the number of pixels in $\Gamma$, which is chosen from a bank of parametrized frequency-time tracks, and each track is referred to as a ``template.'' 

To modify the above algorithm to perform an all-sky search \cite{stochsky,stochtrack_cbc,stochtrack_ecbc}, we use a ``complex signal-to-noise ratio'':
\begin{equation}\label{eq:german_p}
  \mathfrak{p}(t;f) = 
    \frac{2 \sqrt{2}}{\cal N} \left[
    \frac{\tilde{s}_I^*(t;f) \tilde{s}_J(t;f)}{\sqrt{P'_I(t;f) P'_J(t;f)}}
    \right] .
\end{equation}
This statistic preserves the complex phase information, which encodes the direction of the source. As a proxy for the sky location, which is unknown, we add an additional variable $\Delta\tau$ which corresponds to the time delay between the detectors \cite{stochsky}. 
Therefore, we rewrite Eq.~\ref{eq:sum} as
\begin{equation}\label{eq:german_sum}
  \text{SNR}_\text{tot} \equiv
  \frac{1}{N^{1/2}}
  \sum_{\left\{t;f\right\}\in\Gamma} 
  \text{Re}\left[
    e^{2\pi i f \Delta\tau} \mathfrak{p}(t;f) 
    \right],
\end{equation}
and this sum is carried out for many randomly selected clusters $\Gamma$. 

\subsection{Parameterizations}
In any seedless clustering algorithm, $\Gamma$ is chosen such that it is sensitive to the morphology of the the gravitational waves being searched for. There are two types we will consider in this paper, although the method is generic enough to work for any parameterization.

{\em B\'ezier curves.}
For generic narrow-band long transient gravitational waves~\cite{stochtrack,stochsky}, $\Gamma$ is chosen randomly from the set of quadratic B\'ezier curves~\cite{bezier} subject to the constraint that the curve persists for a duration $t_\text{min}$.
Three time-frequency control points determine the template: $P_0$ $(t_\text{start}, f_\text{start})$, $P_1$ $(t_\text{mid}, f_\text{mid})$, and $P_2$ $(t_\text{end}, f_\text{end})$, and the curve is parameterized by $\xi=[0,1]$:
\begin{equation}
  \left(\begin{array}{c} t(\xi) \\ f(\xi) \end{array}\right) =
  (1-\xi)^2 P_0 + 2(1-\xi)\xi P_1 +
  \xi^2 P_2 .
\end{equation}
These arrays allow the sum in Eq.~\ref{eq:sum} to be computed for a large number of templates in parallel.
For practical applications, the number of templates $T$ is typically chosen to be $T={\cal O}(10^4--10^8)$.
To perform a computationally feasible all-sky analysis, $T=2\times10^4$ templates are feasible, and we use this number in the analysis that follows.

{\em Post-Newtonian templates for compact binary coalescences.}
Another parameterization for $\Gamma$ currently in the literature creates templates based on a post-Newtonian model for chirp-like signals created by circular compact binary coalescences \cite{stochtrack_cbc}. 
For searches for compact binary coalescences with seedless clustering, we can use a more specialized template bank consisting of parametrized chirps:
\begin{equation}\label{eq:f_of_t}
  f(t) = \frac{1}{2 \pi} \frac{c^{3}}{4 G M_\text{total}} \sum_{k=0}^7 p_k \tau^{-(3+k)/8} ,
\end{equation}
where
\begin{equation}
  \tau = \frac{\eta c^{3} (t_c -t)}{5 G M_\text{total}},
\end{equation}
where the expansion coefficients $p_k$ can be found in \cite{PN}, $G$ is the gravitational constant and $M_\text{total}$ is the total mass of the binary.
These templates are parameterized by the coalescence time and the chirp mass.
It was shown in~\cite{stochtrack_cbc} that while the waveform depends on the individual component masses, the main features of the signal can be well-approximated by only the chirp mass, and we can approximate that the individual component masses are equal. 

\subsection{Semi-analytical approximation}
We now describe an semi-analytical approximation to the background of our seedless clustering algorithms. 
Seedless clustering, which computes the sum of pixels in a track, divided by the square root of the number of pixels in the track, lends itself to modeling due to its simplicity. By the central limit theorem, the sum of a sufficiently large number of independent random variables, each with a well-defined expected value and well-defined variance, will be approximately normally distributed. Therefore, we expect that the sum of many pixels will approach a normal distribution, given by
\begin{equation}
\textrm{PDF}_{\textrm{Normal}}(z) = \frac{1}{\sigma \sqrt{2 \pi}} e^{-\left(z-\mu\right)^2 /2 \sigma^2}
\label{eq:normal}
\end{equation}


Because seedless clustering measures the maximum $\text{SNR}_\text{tot}$ of many tracks, here we seek the extreme value distribution for $\text{SNR}_\text{tot}$. This is motivated by the desire for a distribution with which to compare those measured from an analysis using the algorithm.
We can analytically compute a probability distribution for this maximum value as follows. 
Given a random sample of $\text{SNR}_\text{tot}$s $(X_1,...,X_N)$, from a continuous distribution with a probability density function $f(x)$ and cumulative density function $F(x)$, the cumulative density function of the maximum of $\text{SNR}_\text{tot}$ is then given by
\begin{equation}
\begin{split}
\textrm{CDF}_{\text{Max}[\text{SNR}_\text{tot}]}(z) = P(\textrm{max}(X_i)<z) \\ 
= P(X_1<z,...,X_N<z) = P(X_1<z)...P(X_N<z)
\end{split}
\end{equation}
where the third equality assumes that the random samples are independent. 
We will discuss the effect of this assumption in section~\ref{sec:CBC}.
In the case where the probability density functions are identical, the equation becomes
\begin{equation}
\begin{split}
\textrm{CDF}_{\text{Max}[\text{SNR}_\text{tot}]}(z) = [P(X < z)]^N = [F_X (z)]^N.
\end{split}
\label{eq:CDF}
\end{equation}
We show below that we can use this equation, where the CDFs are given by Gaussian CDFs, to approximate the seedless clustering distributions.
Even though in our case $F_X (z)$ is derived from a Gaussian distribution, equation~\ref{eq:CDF} is true for any general distribution represented by $F_X (z)$. Hence in cases where the analytic expression for $F_X (z)$ is difficult to derive or approximate, one can use the observed distribution.

\section{Background Study}\label{results}
We can test the approximations by performing the analysis on Monte Carlo Gaussian noise and initial LIGO noise from the Hanford, WA (H1) and Livingston, LA (L1) detectors. We create complex signal-to-noise ratio spectrograms $\mathfrak{p}(t;f)$ using Eq.~\ref{eq:german_p} and analyze each with the various seedless clustering algorithms. 
Following~\cite{stochtrack,stochsky}, we create $\unit[250]{s}$ maps in a band between $100$--$\unit[250]{Hz}$ with spectrogram resolution of $\unit[1]{s}\times\unit[1]{Hz}$ using $50\%$-overlapping Hann windows.
The results for each are as follows.

\subsection{B\'ezier parameterization}
We begin by analyzing the performance of the analytic model on the B\'ezier parameterization.
\begin{figure*}[t]
 \includegraphics[width=3.5in]{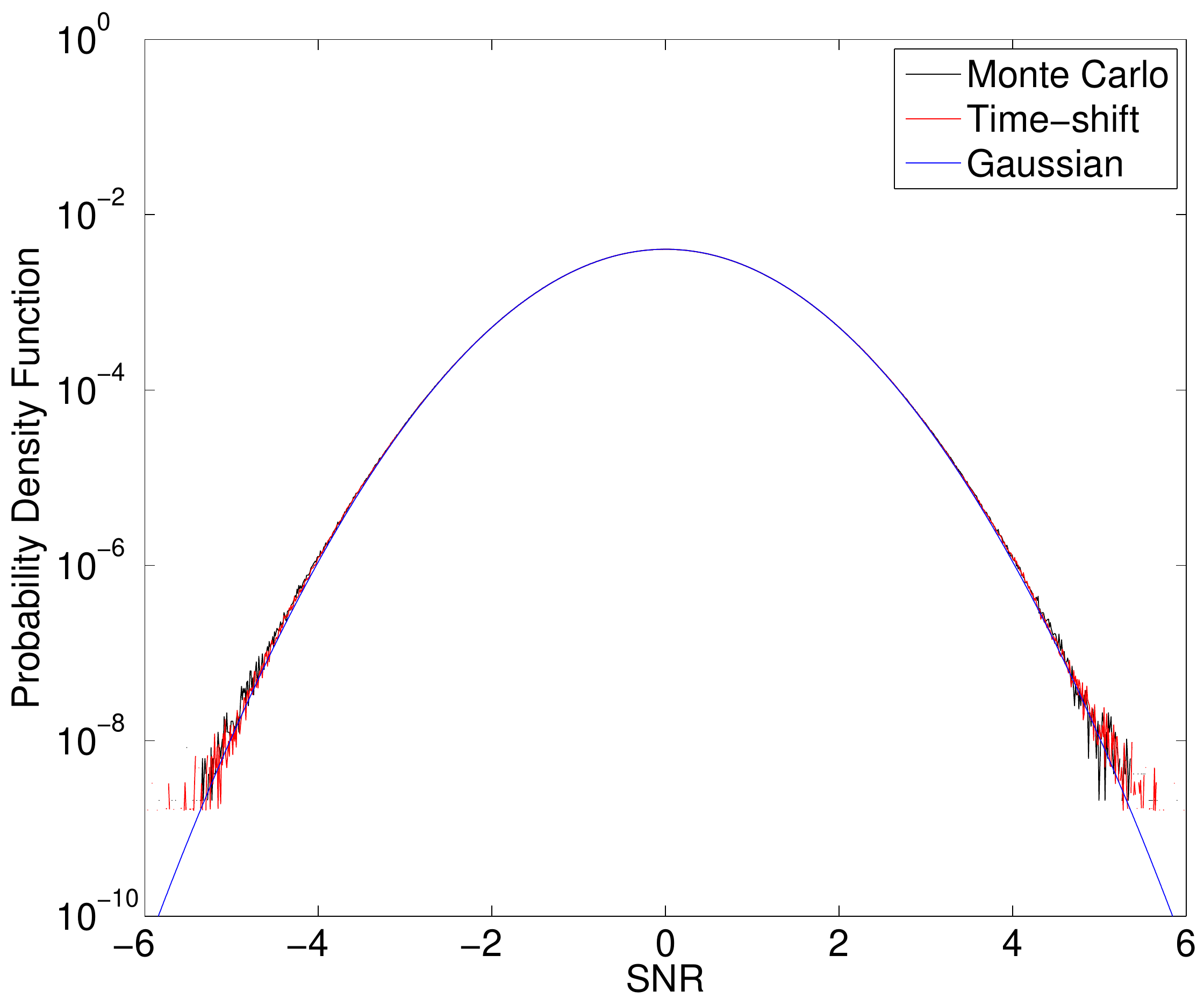}
 \includegraphics[width=3.5in]{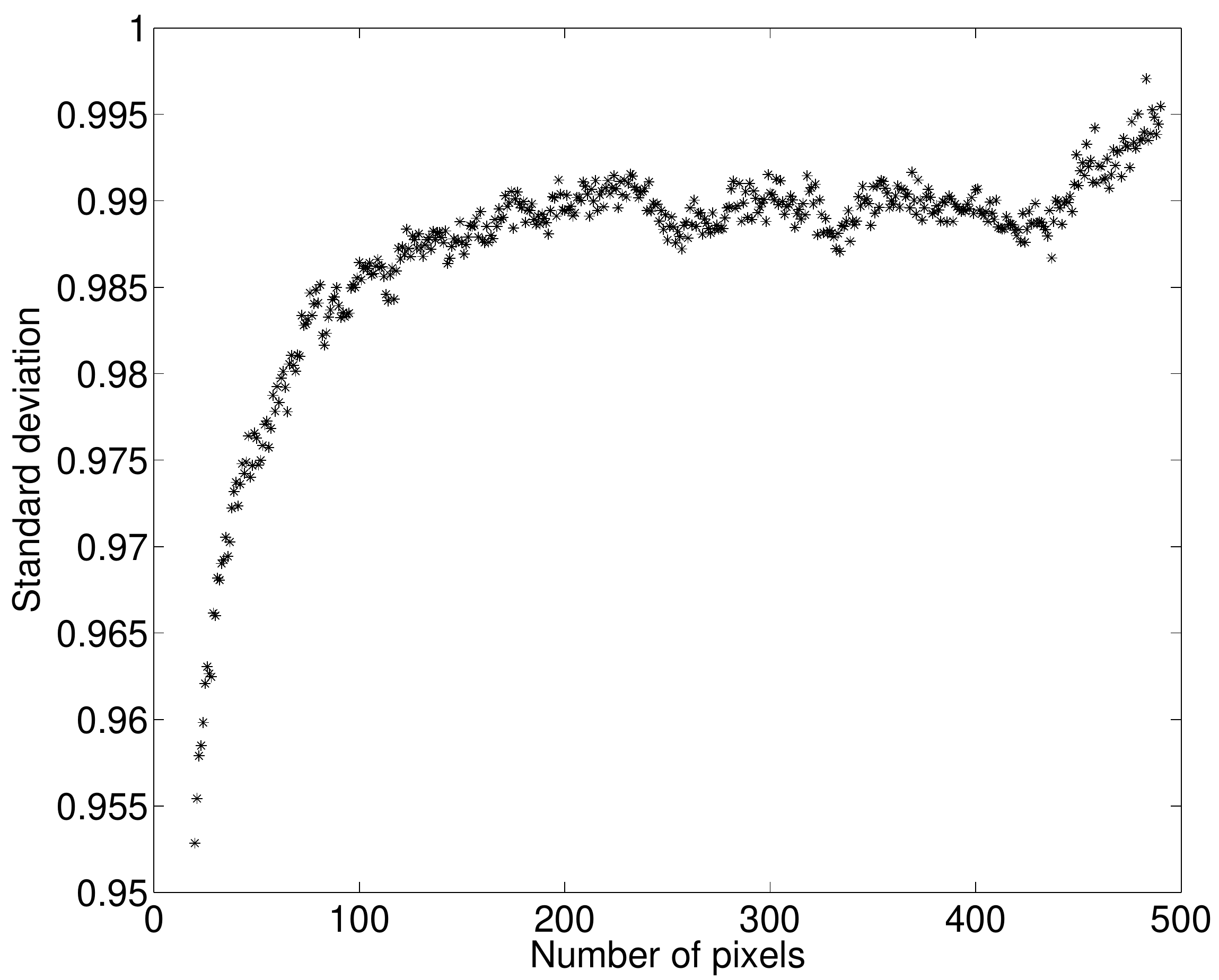}
  \caption{
   The plot on the left is the background distribution for the seedless clustering algorithm cluster SNR defined in equation~\ref{eq:german_sum}.
   Monte Carlo denotes Gaussian colored noise.
   Time-shift denotes time-shifted data with vetoes to limit the effects of instrumental artifacts.   The theoretical line corresponds to the Gaussian approximation to the distribution given by Eq.~\ref{eq:normal}.
   The plot on the right is the standard deviation of $\text{SNR}_\text{tot}$ as a function of track length. 
   The standard deviation differs on the order of a few percent across the track lengths considered. 
   In our analysis, we approximate the standard deviation of $\text{SNR}_\text{tot}$ across track length as constant.
 }
 \label{fig:Hist}
\end{figure*}

We run the seedless clustering algorithm over hundreds of these maps and save $\text{SNR}_\text{tot}$ for each of the tracks in the map.
The left of Fig~\ref{fig:Hist} shows a histogram of the resulting $\text{SNR}_\text{tot}$s for both the Monte Carlo and initial LIGO data. We fit Eq.~\ref{eq:normal} to the resulting distributions. 
We find best fits of $\mu=0.0007$ and $\sigma=0.99$.
The fact that the distribution has approximately a mean of zero and a standard deviation of 1 is expected based on the fact that $\rho$ has a mean of 0 and we use the $\sqrt{N}$ normalization in the $\text{SNR}_\text{tot}$ calculation. 
We find that the agreement is reasonable out to the tails of the distribution.
The right of figure~\ref{fig:Hist} shows the standard deviation of the distribution as a function of track length. The standard deviation differs on the order of a few percent across the track lengths considered. For the sake of simplicity, we assume that the distribution is approximately independent of track length.

We now simulate an all-sky search by performing 100 time-slides in a week of data. The data are processed with a glitch identification~\cite{stamp_glitch} cut as if it were a real analysis. In order to apply the algorithm from~\cite{stamp_glitch}, we assume that the source is optimally oriented with an optimal sky position. To compare this to the analytic approximation in equation~\ref{eq:CDF}, we use the Gaussian fit shown in Fig~\ref{fig:Hist} to approximate the $\text{SNR}_\text{tot}$ distribution. The steps required to turn the $\text{SNR}_\text{tot}$ distribution into a p-value vs.\,SNR distribution are as follows. 
To generate a $\text{SNR}_\text{tot}$ value for a single simulated ``map,'' we generate N random numbers consistent with the Gaussian distribution of mean and variance as estimated above. We then take the maximum value of these values to compute the $\text{Max}[\text{SNR}_\text{tot}]$. To generate a p-value vs.\,SNR distribution, $\text{Max}[\text{SNR}_\text{tot}]$ is generated for M instances, where M is the number of instances required. $\text{Max}[\text{SNR}_\text{tot}]$ is placed in ascending order. The p-value is calculated as an array between 1/M and 1 with spacing given by 1/M. 
For a Gaussian distribution where the mean and standard deviation are the same across all trials, this process can also be performed analytically by simply computing equation~\ref{eq:CDF} for the measured distribution.

We perform two search simulations using the B\'ezier parameterization. The first uses B\'ezier templates computed for a specific search direction. The second loops over time-delays for each template. 
By searching over 40 different time-delays, corresponding to 40 different sky rings, a computationally efficient all-sky search can be performed. This was demonstrated in \cite{stochtrack_cbc} to be sufficient to recover signals in arbitrary directions. This involves a rotation in the complex plane of the individual pixels that make up the tracks. This creates difficulty for the analytic analysis. Because the analysis amounts to a rotation, the 40 time-delays do not correspond to 40 independent trials (which would simply multiply the number of tracks by 40). 
One possibility would be to measure the covariances between the rotated pixels. This situation is similar to \cite{PhysRevLett.107.271102}. In this work, the authors place limits on gravitational-wave strain from different portions of the sky. This was difficult because the distribution of maximum SNR for a sky map contains non-zero covariances that exist between different pixels (or patches) on the sky. They simulate the covariance between pixels numerically, by simulating many realizations that have expected covariances described by the Fisher matrix. In this case, we could numerically compute a covariance matrix, which we can diagonalize to create a basis of non-covariant variables. Then, one would generate random realizations of these non-covariant variables and use the covariance matrix to convert them into a set of randomly generated covariant variables. One difficulty is that the distribution of the non-covariant variables might not be the same as the covariant variables. With this method, we could determine the set of covariant variables which describe the distribution of $\text{SNR}_\text{tot}$.

\begin{figure*}[t]
 \includegraphics[width=3.5in]{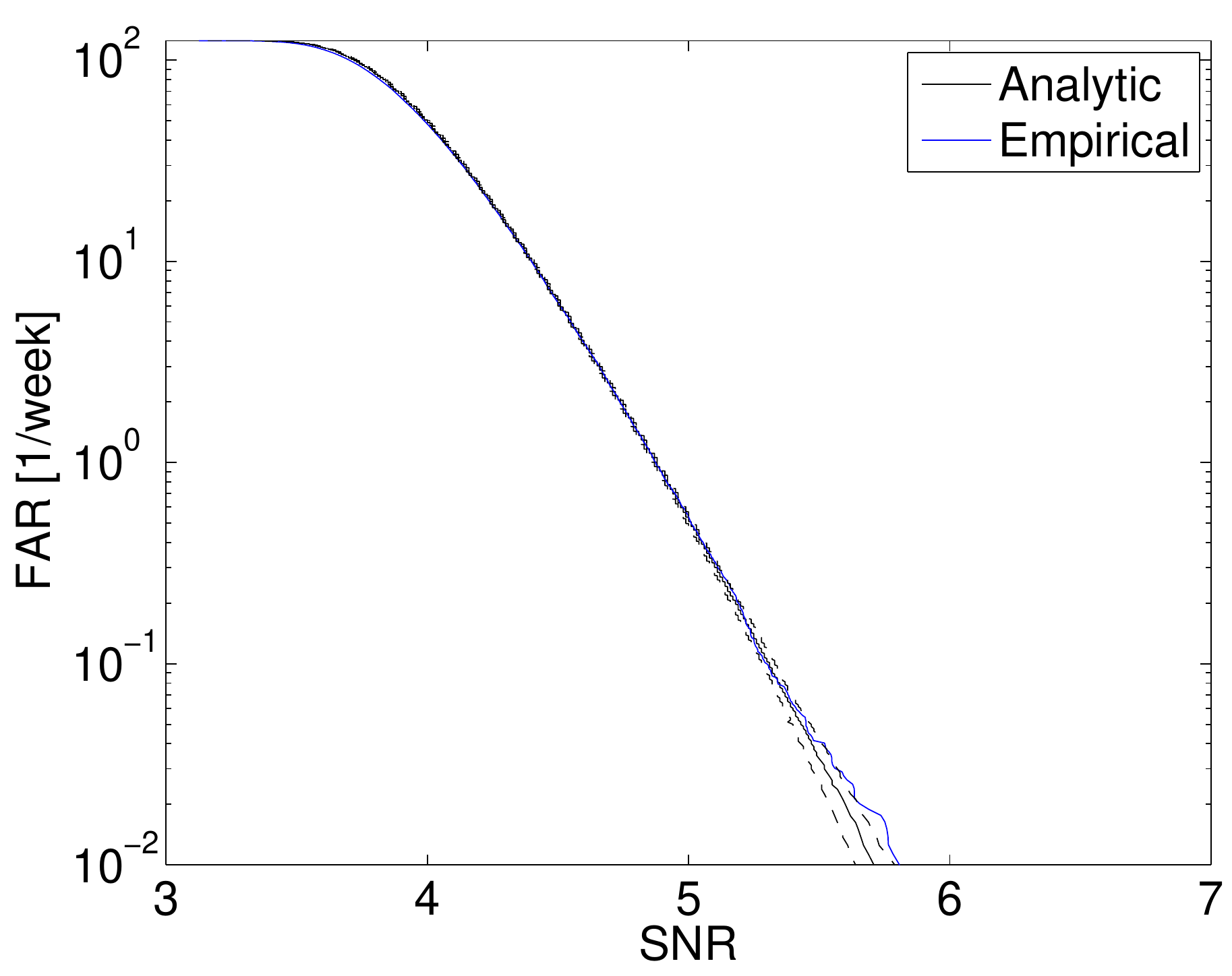}
 \includegraphics[width=3.5in]{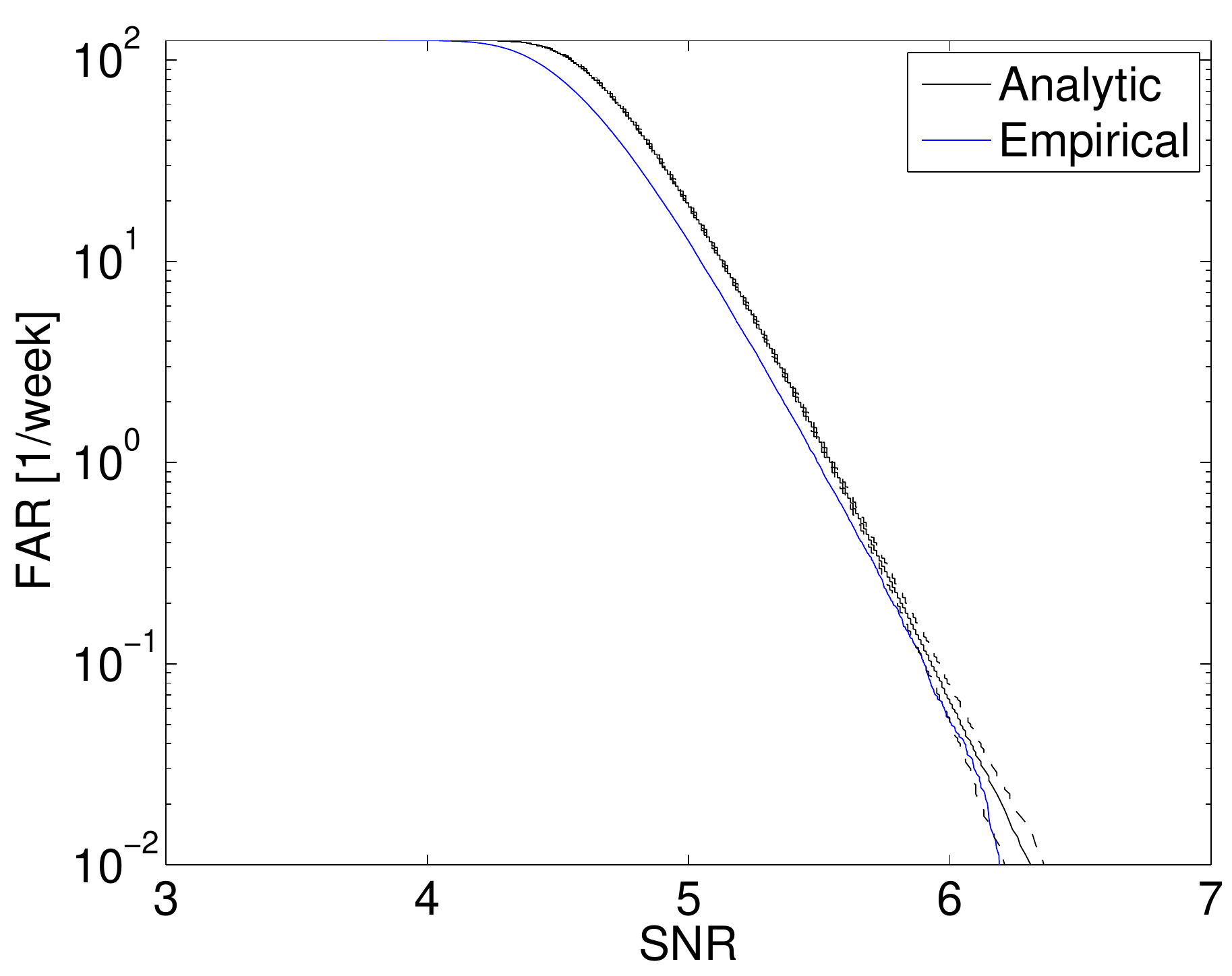}\\
 \includegraphics[width=3.5in]{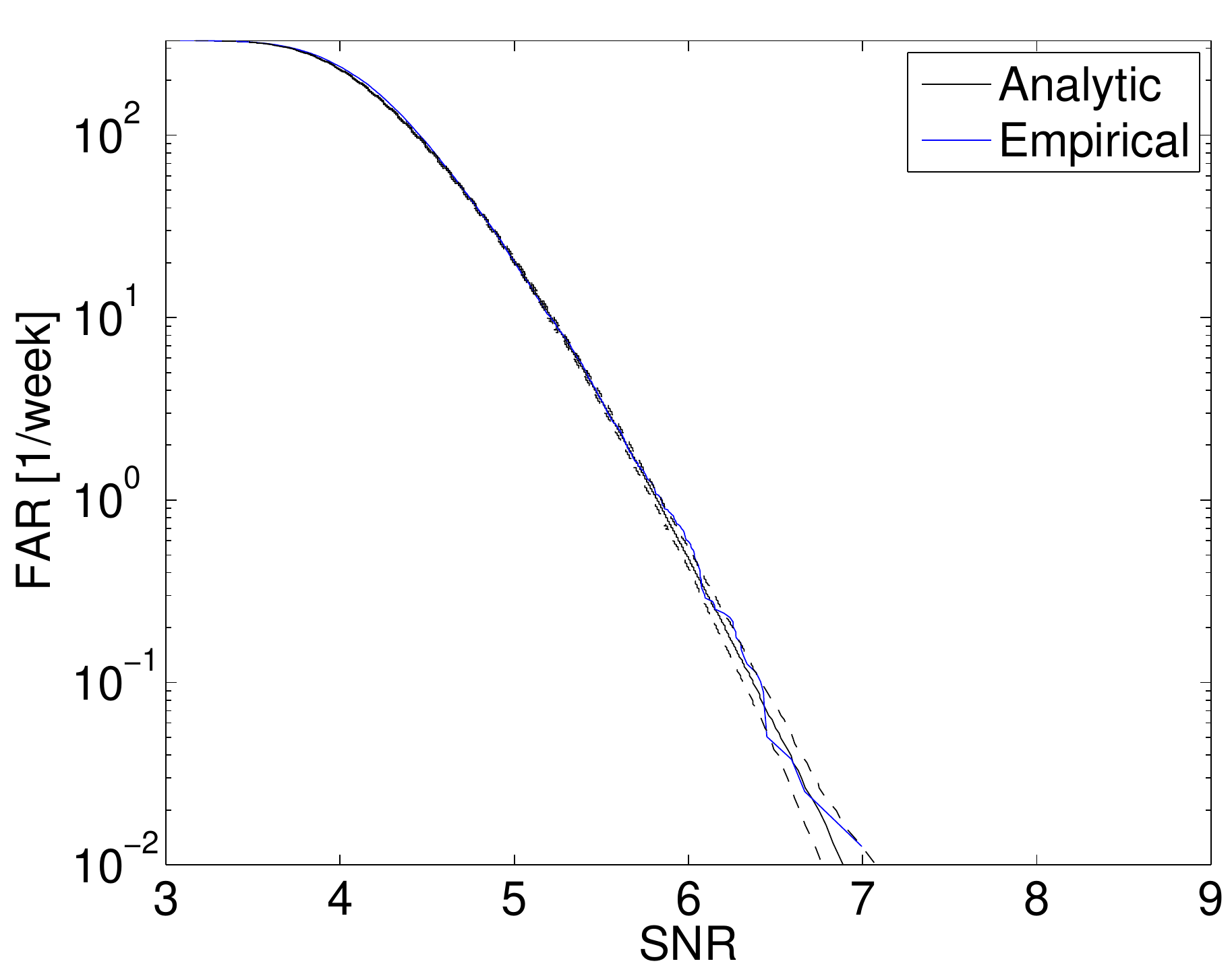}
 \includegraphics[width=3.5in]{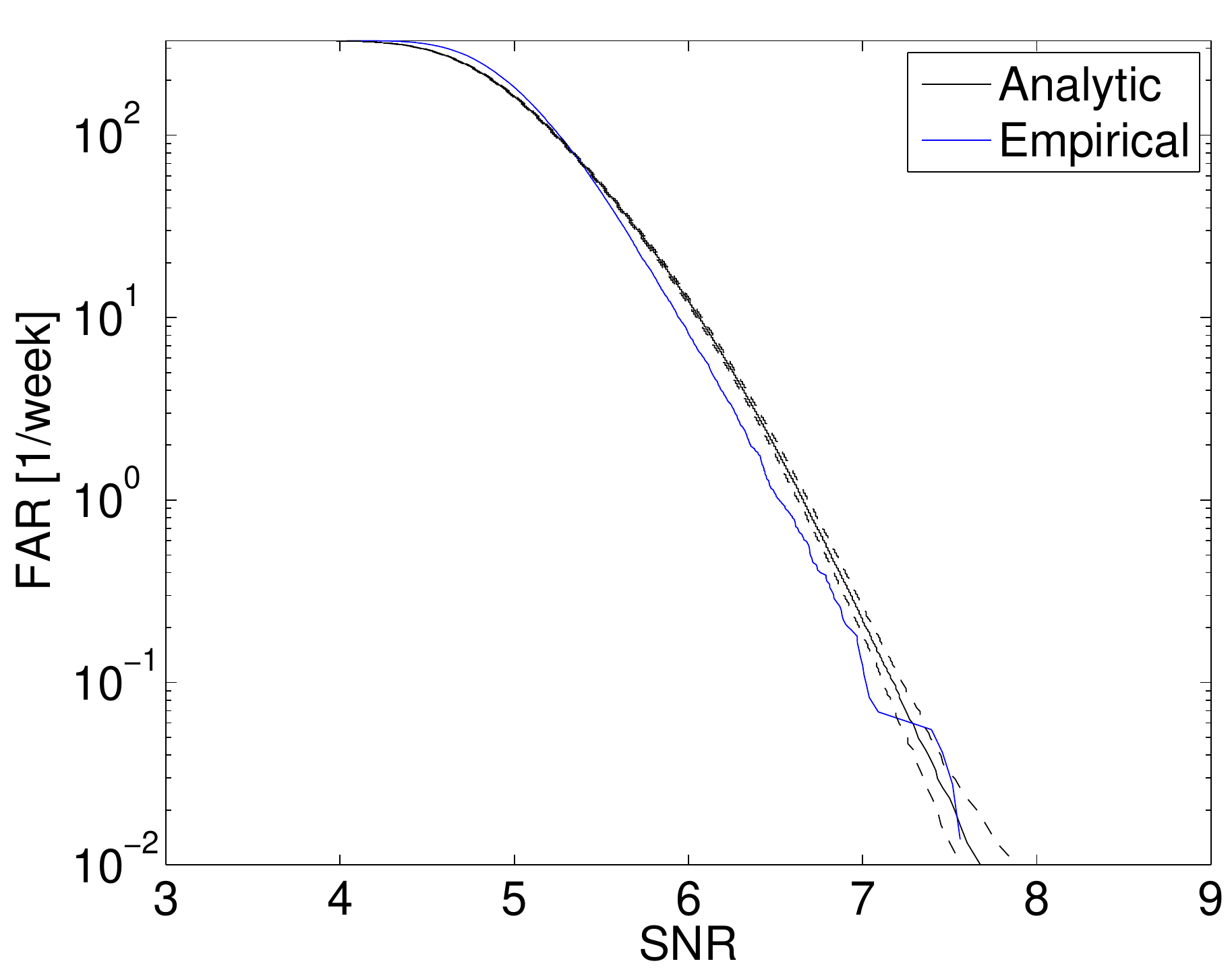}
  \caption{
   Background distributions computed for the seedless clustering algorithm using both B\'ezier and chirp-like templates. 
   The distributions are generated from time-shifted initial LIGO data.
   The top row corresponds to the B\'ezier templates and the bottom row chirp-like templates. 
   The left column corresponds to a directed search (in a specific sky direction) and the right to an all-sky search performed looping over $40$ time-delays for each template.
   The theoretical line corresponds to using Gaussian distributions with standard deviations presented in figure~\ref{fig:STDBezierCBC}.
   The dotted lines correspond to 1\,$\sigma$ error bars on the analytic approximation.
   All analytical background distributions are consistent with the measured background (within $1\,\sigma$).
   }
 \label{fig:Background}
\end{figure*}

The top left of Fig~\ref{fig:Background} demonstrates the analysis for the first simulation using 20,000 tracks, showing both empirical time slides as well as the theoretical approximation method explained as before (both the 10th, 50th, and 90th percentiles). We find excellent agreement between the analytic model and empirical time-slides. There are a number of reasons we might expect small disagreements between the theoretical model and the empirical results. First of all, generating purely random numbers to approximate $\text{SNR}_\text{tot}$ is an approximation. This is because the tracks are analyzed on the same map, and therefore overlapping tracks will have correlated $\text{SNR}_\text{tot}$ values. Another implicit assumption is that the pixels in the tracks are uncorrelated. This assumes that the noise is Gaussian and stationary and ignores the correlation between pixels in the maps. However, the cross-power statistic uses PSD's from adjacent pixels (in the time direction) to estimate $\sigma$ (from Eq.~(\ref{eq:sum})). This is to avoid a bias in pixel SNR for an isolated loud pixel. This means there is a correlation between adjacent pixels. A similar concern arises from the fact that real detectors have noise transients and non-stationary noise, which violate some of the approximations used here. 
The top right of Fig~\ref{fig:Background} demonstrates the analysis for the second simulation using 20,000 tracks as well. In this case, we multiply the number of templates by 40, corresponding to the 40 time-delays. We expect minor disagreement, which is present, due to the conservative assumption that each loop corresponds to an independent trial. We show the $\text{Max}[\text{SNR}_\text{tot}]$s required for a 5-sigma gravitational-wave detection using the B\'ezier parameterization in figure~\ref{fig:STDBezierCBC}.

\subsection{Compact binary coalescence parameterization}
\label{sec:CBC}
We now analyze the performance of the analytical model on the chirp templates.
We create maps assuming Gaussian noise consistent with the design sensitivity of Advanced LIGO.
Following~\cite{stochtrack_cbc}, we create $\unit[660]{s}$ maps in a band between $10$--$\unit[150]{Hz}$ with a spectrogram resolution of $\unit[1]{s}\times\unit[1]{Hz}$.

We perform a similar analysis to the above. 
We find best fits for the Gaussian distribution of $\mu=-0.004$ and $\sigma=1.06$.
The major difference between the B\'ezier and chirp-like template analysis is the degree of correlations between the drawn tracks. In the B\'ezier case, the tracks are drawn randomly and the degree of correlation is simply determined by the overlap between the tracks. In the chirp-like template case, the degree of correlation is much higher, despite the significantly fewer templates used in the analysis. This correlation arises from the step in time and overlap in parameter space between chirp-like templates of similar chirp mass.

\begin{figure*}[t]
 \includegraphics[width=3.5in]{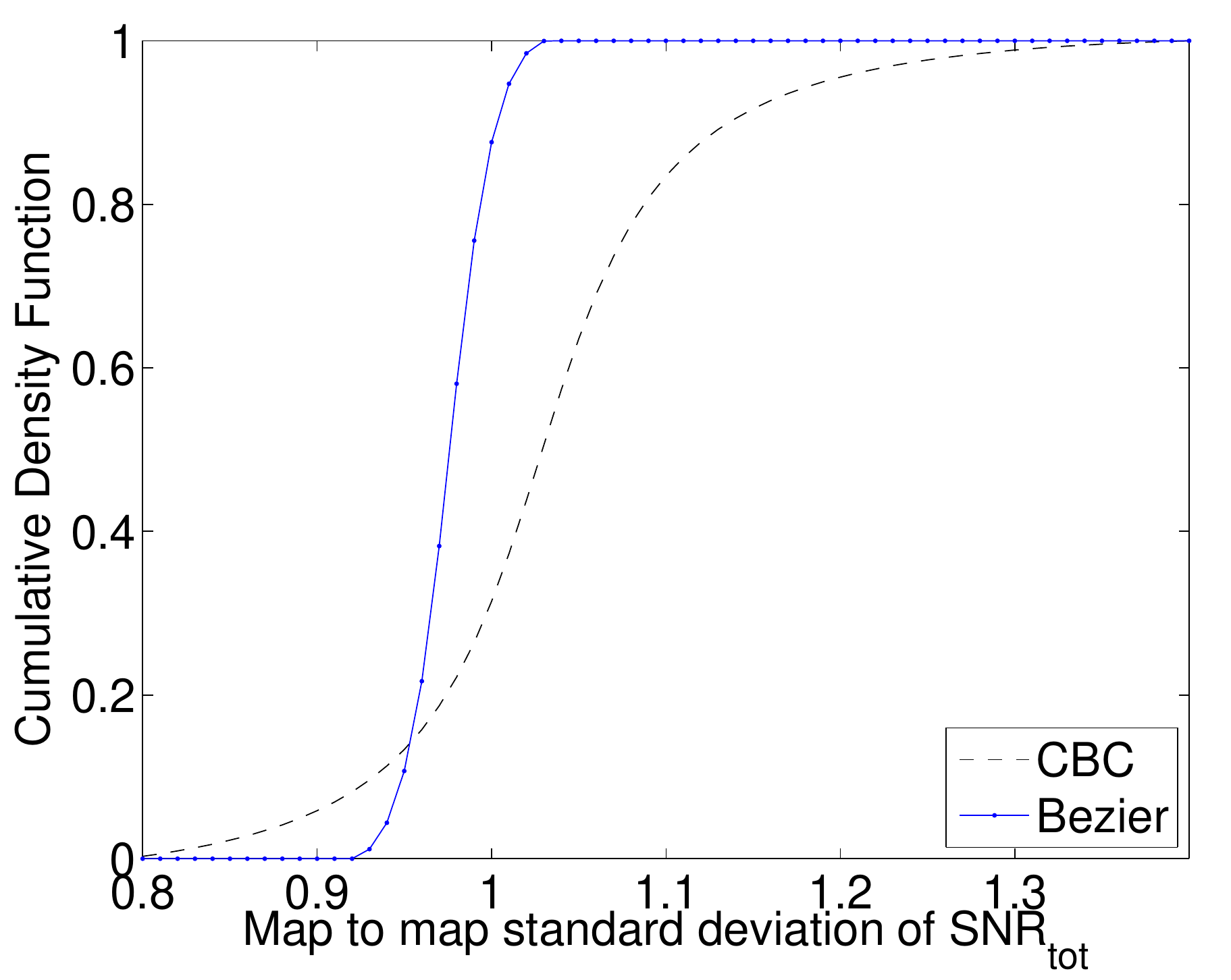}
 \includegraphics[width=3.5in]{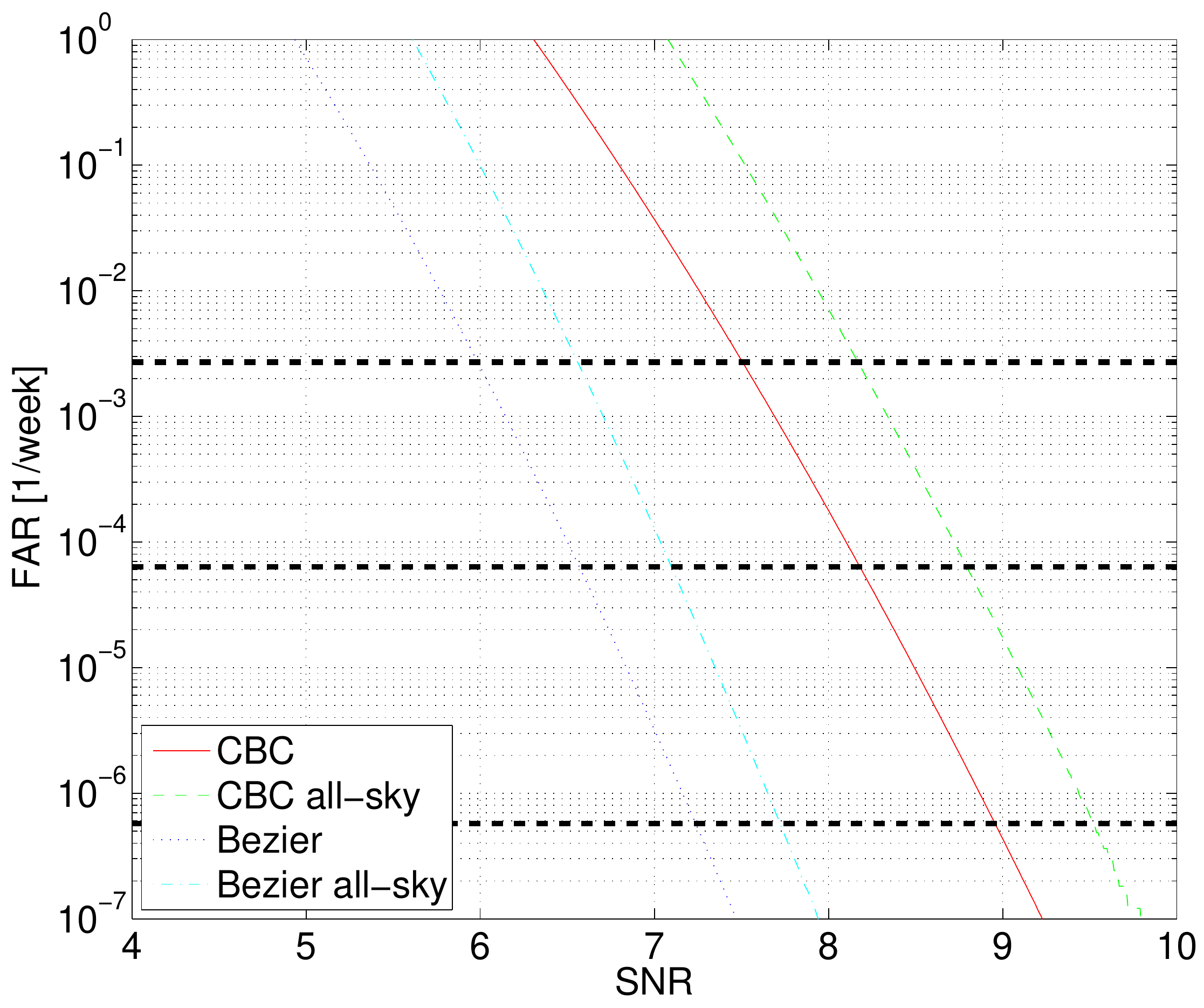}
  \caption{
   The plot on the left is the cumulative density function of the standard deviation of the $\text{SNR}_\text{tot}$ for the two parameterizations. 
   Distribution is significantly broader for chirp-like templates due to track correlations.
   The plot on the right is the background distribution using the analytic approximation for the cases considered here.
   It shows the $\text{Max}[\text{SNR}_\text{tot}]$s required for a $3 \sigma$, $4 \sigma$, and $5 \sigma$ gravitational-wave detection using seedless clustering in dotted horizontal lines. 
  }
 \label{fig:STDBezierCBC}
\end{figure*}

This correlation changes the standard deviation of the $\text{SNR}_\text{tot}$ of the tracks in individual maps. Fig~\ref{fig:STDBezierCBC} demonstrates the cumulative density function of the standard deviation of the $\text{SNR}_\text{tot}$ for the two parameterizations. 
In the B\'ezier case, the standard deviation of $\text{SNR}_\text{tot}$ is approximately a step function, which allows for the use of a single standard deviation to cover all cases.
The distribution is significantly broader for chirp-like templates due to track correlations. 
It is for this reason that we modify the B\'ezier p-value algorithm by drawing from the measured distribution of standard deviations of the maps when drawing from the Gaussian distribution. 
The bottom of Fig~\ref{fig:Background} demonstrates the algorithm using both empirical Monte carlo noise as well as the theoretical approximation method explained as before (both the 10th, 50th, and 90th percentiles) for both a directed and all-sky search.
We show the $\text{Max}[\text{SNR}_\text{tot}]$s required for a 5-sigma gravitational-wave detection using the chirp-like parameterization in figure~\ref{fig:STDBezierCBC}.

\section{Conclusions and Future Work}\label{conclusions}
In previous work, we showed how a seedless clustering algorithm could significantly improve the sensitivity of searches for long-lived, unmodeled gravitational-wave transients~\cite{stochtrack,stochsky,stochtrack_cbc,stochtrack_ecbc}.
Here, we show how the simplicity of the search statistic allows for the development of a semi-analytic approximation to the background generated by the algorithm and compared the performance using a week of LIGO S5 time-shifted data.
We described algorithmic subtleties not addressed by this model and quantify the errors between the model and the measured distribution. We argued that it will be useful for pipeline characterization, as well as potentially for low-latency FAP reporting for gravitational-wave searches.

In the future, we will move beyond the simple models presented here to more complicated models. Some examples could be using non-Gaussian distributions, such as the Student-t distribution, to better approximate the tails of the distribution, which is where we expect the strongest disagreement. Other ideas include using the Edgeworth expansion to put bounds on the deviation from Gaussianity. As the tracks in individual maps are correlated (due to the fact that some will overlap and use the same pixels), we could also consider generating correlated random values when generating our distributions for $\text{SNR}_\text{tot}$.

\begin{acknowledgments}
MC is supported by National Science Foundation Graduate Research Fellowship Program, under NSF grant number DGE 1144152.
ET was a member of the LIGO Laboratory, supported by funding from United States National Science Foundation.
NC work was supported by NSF grant PHY-1204371.
LIGO was constructed by the California Institute of Technology and Massachusetts Institute of Technology with funding from the National Science Foundation and operates under cooperative agreement PHY-0757058.
This paper has been assigned document number LIGO-P1500053.
\end{acknowledgments}

\bibliography{pvalues}

\end{document}